# Exchange-driven spin relaxation in ferromagnet/oxide/semiconductor heterostructures

Yu-Sheng Ou[1], Yi-Hsin Chiu[1], N. J. Harmon[2], Patrick Odenthal[3], Matthew Sheffield[1], Michael Chilcote[1], R. K. Kawakami[1,3], M. E. Flatté[2], E. Johnston-Halperin[1,†]

[1]*Department of Physics, The Ohio State University, Columbus, OH 43210-1117, USA*
[2]*Department of Physics & Astronomy, The University of Iowa, Iowa City, IA 52242-1479, USA*
[3]*Department of Physics & Astronomy, University of California, Riverside, CA 92521, USA*

Abstract

We investigate electron spin relaxation in GaAs in the proximity of a Fe/MgO layer using spin-resolved optical pump-probe spectroscopy, revealing a strong dependence of the spin relaxation time on the strength of an exchange-driven hyperfine field. The temperature dependence of this effect reveals a strong correlation with carrier freeze out, implying that at low temperatures the free carrier spin lifetime is dominated by inhomogeneity in the local hyperfine field due to carrier localization. This result resolves a long-standing and contentious question of the origin of the spin relaxation in GaAs at low temperature when a magnetic field is present. Further, this improved fundamental understanding paves the way for future experiments exploring the time-dependent exchange interaction at the ferromagnet/semiconductor interface and its impact on spin dissipation and transport in the regime of dynamically-driven spin pumping.

Gallium arsenide has a long history as a canonical test bed for the investigation of fundamental spin relaxation properties [1-3] and the development of prototype spintronic structures based on ferromagnet (FM)/GaAs heterostructures [4-6]. However, despite its long history, there remain significant questions regarding the fundamental spin relaxation/dissipation processes in the GaAs spin channel itself. Specifically, spin-phonon coupling [7], energy dependence of the Lande g-tensor [8], and inhomogeneities in the hyperfine interaction [9] have all been proposed to explain the low-temperature ($< 50$ K) spin relaxation in the presence of an applied magnetic field. This absence of clarity prevents the continued use of GaAs based heterostructures to explore emerging areas of current interest. For example, it would be natural to use FM/GaAs heterostructures to elucidate the dynamic exchange mechanisms underlying the recent development of novel ferromagnetic resonance (FMR) [10-13] and thermally-driven spin injection processes [14-18]. Indeed, the ability of ultrafast pump-probe spectroscopies to probe GaAs spin dynamics directly in the time domain [2] would in principle allow for a direct measurement of the dynamic exchange coupling and dissipation at FM/GaAs interfaces if we were able to more completely understand the underlying interactions.

Here we present a systematic investigation of the free carrier spin lifetime in Fe/MgO/GaAs heterostructures and bare GaAs films that identifies inhomogeneities in the hyperfine interaction due to the random distribution of Si donors, as proposed in Ref. [9], as the limiting mechanism in determining the spin relaxation rate in this critical experimental regime. By examining Fe/MgO/GaAs heterostructures, we are able to access large effective nuclear fields due to the exchange-driven hyperfine coupling at low applied field ($< 3$kG). Comparable nuclear fields in bare GaAs require applied fields on the order of 10s of kG. This ability to tune the nuclear field using exchange coupling allows us to demonstrate the importance of inhomogeneous nuclear fields in governing electron spin relaxation in both bare GaAs films and FM/GaAs heterostructures. Thus we resolve a long-standing and contentious question of the origin of the spin lifetime in low temperature GaAs when a magnetic field is present. This more complete understanding in turn allows a quantitative description of the dynamic, exchange mediated, electron-nuclear interactions in our FM/GaAs nanostructures.

Our theory of the inhomogeneous hyperfine interaction is depicted in left panel of Fig. 1 (a). Large nuclear fields are induced by the process of dynamic nuclear polarization [1] from a non-equilibrium electron spin polarization generated via a combination of optical excitation and exchange coupling to the proximal FM. Since the electron-nuclear spin transfer is most efficient in the vicinity of neutral Si donors (green circles), the magnitudes of the nuclear fields (red arrows) within the GaAs are strongly inhomogeneous due to carrier localization (yellow circles). Transitions between adjacent field environments cause transverse spin relaxation [9].

The schematic structure of the samples studied in this work is shown in Fig. 1 (b), with layer thicknesses of: 8 nm MgO/10 nm Fe/0.2 nm MgO/120 nm Si doped n-GaAs ($7 \times 10^{16}$/cm$^3$)/400 nm In$_{0.5}$Ga$_{0.5}$P/n$^+$-GaAs (100) substrate. These samples were synthesized according to Ref. [20] with the thickness of the MgO layer optimized to maximize the exchange coupling at the Fe/GaAs interface [20]. Fig. 1 (c) shows the simulated band structure of the sample calculated using a self-consistent one-dimensional Schrödinger/Poisson solver (BandEng). The band offset at the interface (Fig. 1 (c), inset) is determined by previous studies using x-ray and ultraviolet photoelectron spectroscopies to study the band structure of Fe/MgO/GaAs tunnel junction [21]. A control sample is grown with a similar structure but without the Fe/MgO layer, and both samples are mounted face-down on 100 μm thick sapphire wafers so that the n$^+$-GaAs substrates can be removed by selective wet etching using the $In_{0.5}Ga_{0.5}P$ layer as a chemically-selective etch stop [22].

We explore the strength of the interfacial exchange interaction in these heterostructures via time-resolved Kerr rotation (TRKR) to evaluate the interface quality [19,20,23]. A schematic of the technique is shown in Fig. 1(b); a circularly-polarized (CP) pump pulse excites electron spins in GaAs along its propagation direction, which then precess in the presence of a transverse magnetic field ($B_{tot}$). After a time delay $\Delta t$, the Kerr rotation ($\theta_K$) of a much weaker linearly-polarized (LP) probe pulse measures the spin component along its propagation direction. The Kerr rotation time trace ($\theta_K$ vs. $\Delta t$) reveals the temporal evolution of the photoexcited electron spins and can be described by the following equation [24]:

$$\theta_K(\Delta t) = \theta_0 (e^{-(\Delta t/T_2^*)} + N_0 e^{-(\Delta t/T_h)}) \cos(\omega_L \Delta t + \phi) \quad (1)$$

where $\theta_0$ is the maximal Kerr angle and $N_0$ is the ratio of photoexcited to equilibrium carriers at $\Delta t = 0$, $T_2^*$ is the ensemble transverse electron spin relaxation time, $T_h$ is the hole carrier lifetime, $\omega_L = g\mu_B B_{tot}/\hbar$ is the Larmor precession frequency and $\phi$ is the phase of the spin precession. The two exponential terms reflect the fact that the spin polarization can live much longer than the lifetime of the photoexcited carriers, with the former extending up to nanoseconds and the latter typically less than 100 ps [25]. Independent of lifetime effects, $\omega_L$ provides a local magnetometry that measures both the applied field, $B_{app}$, and any local effective fields, $B_{loc}$, experienced by the photoexcited electron spins, $B_{tot} = B_{app} + B_{loc}$.

The TRKR time scans for both the Fe/MgO/GaAs heterostructure and the GaAs control are shown in Fig. 2 (a). Scans are acquired at a temperature $T = 5$ K and an applied field $B_{app} = 12$ kG. Laser pulses of 130-fs duration and 76 MHz repetition rate are generated by a mode-locked

Ti-Sapphire laser, and are split into pump and probe pulse trains whose power ratio is ~7, with a time-averaged pump power density of 119 W/cm$^2$. The clear difference in $\omega_L$ (or equivalently, $B_{tot}$) between the Fe/MgO/GaAs and GaAs structures implies a variance in $B_{loc}$ between the two samples (roughly -2 kG and +0.2 kG, respectively). The magnitude and sign of $B_{loc}$ in Fe/MgO/GaAs is consistent with previous FPP measurements, and has been attributed to a hyperpolarization of the Ga and As nuclei [19,23].

Figure 2 (b) shows a schematic diagram detailing the fundamental interactions underlying this effect. In the presence of a transverse $\boldsymbol{B_{app}}$, a non-equilibrium electron spin population with a net polarization $\boldsymbol{S_0}$, is excited in the GaAs layer by the circularly polarized optical excitation. From this initial population, we consider separately free carrier spins that reflect from the Fe/MgO layer [26] and spins that evolve purely within the GaAs. The former will acquire a net orientation parallel to the magnetization of the Fe layer through the FPP effect, $\boldsymbol{S_{FPP}}$, while the latter will relax antiparallel to the applied field, $\boldsymbol{S_{rel}}$ (the Lande g-factor in GaAs is -0.44 [27]). These two non-equilibrium electron spin populations both act to dynamically polarize nuclear spins ($\boldsymbol{I}$) via the hyperfine interaction ($H_{hyperfine} = A\boldsymbol{I}\cdot\boldsymbol{S_i}$, where $A$ is the product of nuclear and electron Bohr magneton and the probability density of electron wave function at the nuclear sites and $i = rel$ or $FPP$) and the polarized nuclear spins in turn create an effective local field $\boldsymbol{B_n^{tot}}$ acting on the photoexcited spins, $\boldsymbol{B_n^{tot}} \propto -\boldsymbol{B_{app}}\left(\boldsymbol{B_{app}}\cdot\left(-\boldsymbol{S_{rel}} + \boldsymbol{S_{FPP}}\right)\right)/B_{app}^2$, for $\boldsymbol{B_{app}}$ much larger than the nuclear dipole-dipole field (~10G) and Knight field (~100G) [1,28]. This analysis identifies the local magnetic field identified by the Larmor magnetometry shown in Fig. 2(a), $B_{loc}$, as arising from an effective nuclear field, $B_n^{tot}$, due to the optically induced non-equilibrium nuclear polarization. Further evidence for the nuclear origin of $B_{loc}$ can be found in the resonant suppression of $B_{loc}$ at the various nuclear magnetic resonance (NMR) frequencies of the Ga and As nuclei (see Supplementary Information). It should be noted that since $S_{FPP}$ is antiparallel to $S_{rel}$, the resulting $\boldsymbol{B}_n^{tot}$ can be antiparallel (negative sign) or parallel (positive sign) to $\boldsymbol{B}_{app}$, depending on the competition between the FPP and Zeeman relaxation mechanisms. The observation of $\boldsymbol{B}_n^{tot}$ = -2 kG in Fe/MgO/GaAs indicates that $\boldsymbol{B}_n^{tot}$ is dominated by $S_{FPP}$ while the fact that $\boldsymbol{B}_n^{tot}$ = +0.2 kG in the GaAs control indicates that the nuclear polarization arises from $S_{rel}$.

Although the magnitude and sign of $\boldsymbol{B}_n^{tot}$ in Fe/MgO/GaAs is a strong indication of FPP, more compelling evidence is the ferromagnetic imprinting of the nuclear spin polarization [19,23]. As can be seen in Fig. 2 (c) the dependence of $B_{tot}$ on $B_{app}$ (top panel) has both a linear component (from the Larmor dependence on $B_{app}$) and a component that tracks with the magnetization of the Fe layer, switching at fields below the experimental resolution (~ 0.02 kG) and saturating at $B_{app}$ ~ ±3 kG. This behavior is more clearly seen in the bottom panel of Fig. 2(c) where the linear Zeeman dependence has been subtracted. This is in contrast to the behavior in the GaAs control, where $B_{tot}$ ($\omega_L$) and $B_n^{tot}$ scale linearly with $B_{app}$ (open circles) [1,28,29]. These results are both quantitatively and qualitatively consistent with previous studies [19,23], and confirm the high interfacial quality of the sample.

We now consider the impact of this interfacial exchange coupling and consequent nuclear polarization on the spin relaxation/dissipation in the GaAs layer. Figures 3 (a) and (b) show $T_2^*$ and the magnitude of $\boldsymbol{B}_n^{tot}$ ($|B_n^{tot}|$) as a function of applied field, $\boldsymbol{B}_{app}$, respectively. A remarkable correlation is evident, with $|B_n^{tot}| \sim 1/T_2^*$. There are two distinct regimes evident in these measurements. First, for $B_{app}$ below 0.5 kG (Figs. 3 (c) and (d)) there is a strong enhancement of $T_2^*$ and concurrent suppression of $|B_n^{tot}|$. This is a well-known effect arising from the nuclear depolarization driven by the nuclear dipole-dipole coupling [1,28]. Second, for fields above 0.5 kG there is a competition between the nuclear field generated by the FPP effect, $B_n^{FPP}$, and the nuclear field generated by conventional spin relaxation from Zeeman splitting in the conduction band, $B_n^Z$. As can be seen in Fig. 2 (b), with increasing $B_{app}$ the FPP driven polarization is initially much larger than the Zeeman driven polarization, but saturates as the magnetization saturates at $B_{app} \sim$ 3 kG. In contrast, the Zeeman driven polarization grows slowly but continuously, increasing linearly for the entire field range studied here. Since these two contributions have opposite sign (Fig. 2) their competition gives rise to an inflection point in the total nuclear field, $B_n^{tot} = B_n^{FPP} + B_n^Z$, as can be seen in the maximum in $|B_n^{tot}|$ in Fig. 3 (d).

In general, this correspondence between $|B_n^{tot}|$ and $T_2^*$ strongly indicates that the dominant spin relaxation in this regime is via hyperfine coupling. To gain insight into the origin of this hyperfine-dominated spin relaxation, we consider a theory in which the inhomogeneous nuclear field is due to the non-uniform donor distribution in the GaAs (Fig. 1 (a)), leading to inhomogeneous dephasing of the photoexcited electron spins [9]. In this theory, $S_{FPP}$ and $S_{rel}$ can both relax into donor-bound localized states surrounding the Si dopants in the GaAs as shown in the left panel of Fig. 1 (a). These trapped spins can either directly hyperpolarize nuclei within their Bohr radius (path 1) or polarize donor electrons via the exchange interaction [1,31] that then hyperpolarize surrounding nuclei (path 2), resulting in a puddle of hyperpolarized nuclear spin oriented either parallel (FPP) or anti-parallel (Zeeman) to $B_{app}$. These randomly located polarized nuclei in turn give rise to an inhomogeneous nuclear field distribution that leads to the dephasing of itinerant photoexcited carriers that move across those donor sites. The spin relaxation via path 1 can be calculated using a theory of continuous-time-random-walk for spin [32,33]. As was recently shown in Ref. [9], in the motional narrowing limit the existence of the nuclear field inhomogeneity gives rise to an anisotropic spin relaxation term, $1/T_2^* \sim (B_n^{tot})^2$, which fits our data at $T = 5K$ well, supporting the validity of this interpretation (see Supplementary Information).

Critically, this theory makes two implicit predictions about the expected behavior of $T_2^*$ as a function of the temperature of the sample, $T$, and $\boldsymbol{B}_{app}$. Considering first the effect of the sample temperature, we note that raising the system temperature should weaken the hyperfine coupling due to the thermal activation of localized carriers [1]. This in turn should lead to a more homogeneous nuclear field as well as an overall decrease in $|B_n^{tot}|$ as thermal depolarization of the nuclear bath competes with dynamic nuclear polarization as shown in the right panel of Fig. 1 (a). This decrease in inhomogeneity in the nuclear field should in principle lead to an enhancement of $T_2^*$.

These trends are clearly observed in the temperature dependent data presented in Figs. 4 (a) and (b) for $|B_n^{tot}|$ and $T_2^*$, respectively. Considering first data taken for $\boldsymbol{B}_{app} = 0.18$ kG (black circles) and at temperatures below 40 K, we see a monotonic decrease in $|B_n^{tot}|$ and a monotonic increase in $T_2^*$ for increasing temperature. For temperatures above 40 K, the trend in $|B_n^{tot}|$ continues to monotonically decrease but the increase in $T_2^*$ shows a local maximum, with $T_2^*$ decreasing for higher temperatures. This behavior is qualitatively consistent with a continuous decrease in the strength of hyperfine-induced dephasing of the spin ensemble until it is no longer the dominant spin relaxation mechanism and is quantitatively consistent with the temperature scale for the thermal ionization of the Si dopants (full ionization is expected at roughly 69 K [34]). Comparison of the high temperature behavior of $T_2^*$ with previous reports in bare GaAs [2] suggests that this regime is dominated by D'yaknov-Perel (DP) spin relaxation [2,7] (dashed black line).

We note that this non-monotonic temperature dependence of $T_2^*$ was also observed in bare GaAs, but at much higher $\boldsymbol{B}_{app}$ (> 10 kG) [2,35], and is inconsistent with the recent prediction that spin-phonon coupling is the dominant spin relaxation pathway at low temperature in the presence of a significant $B_{app}$ [7]. The derived spin relaxation rate based on the spin-phonon coupling model is proportional to $(B_{app})^2$ at a fixed temperature, and in the low $B_{app}$ region discussed here, the rate is too small to account for the measured magnitude of $T_2^*$ (see Supplementary Information). A comparison with previous studies [2,35,36] suggests that our theory of spin relaxation via inhomogeneous nuclear fields may also be applied to bare GaAs when a large nuclear field (O (~1 kG)) is present. This can be achieved by optically pumping a non-equilibrium nuclear polarization at large $B_{app}$ (>10 kG)) in our measurement geometry. Our results, both in Fe/MgO/GaAs and bare GaAs samples, clearly identify the peak of $T_2^*$ as the outcome of competition between two spin relaxation mechanisms, inhomogeneous hyperfine interactions and D'yaknov-Perel (DP) spin relaxation.

The second, correlated, prediction of our model is that suppressing the hyperfine coupling should cause the local maximum in $T_2^*$ at 40 K to disappear and allow the next most dominant spin relaxation mechanism (presumably DP in these samples) to be evident at all temperatures. The data in Fig. 3 (c) and (d) provide a path to realizing just such a measurement through the low field dipole-induced depolarization of $|B_n^{tot}|$. Reducing $B_{app}$ from 0.18 kG to 0.10 kG dramatically reduces $|B_n^{tot}|$ from +2 kG to +1 kG at $T$= 5 K, and the data in Fig. 4 (a) show that this suppression persists to higher temperature. This reduction in nuclear spin polarization leads to a suppression of the local maximum in $T_2^*$ at 40 K, and $T_2^*$ converges toward the DP prediction across the entire measured temperature range, as predicted above. The failure to fully recover the DP prediction can be explained by the fact that the finite length of our mechanical delay line and laser repetition rate place a lower bound on the value of $B_{app}$ for which we can experimentally resolve $T_2^*$. As a result, we cannot fully suppress $|B_n^{tot}|$ and therefore must measure in a regime with some residual hyperfine-driven inhomogeneity.

In conclusion, we observe a strong dependence of electron spin relaxation time on the FPP-enhanced hyperfine field in Fe/MgO/GaAs heterostructures. Our results are consistent with a model of inhomogeneous broadening of the effective nuclear field due to carrier localization at Si donors at low temperature, and clarify the origin of a local maximum in the value of $T_2^*$ as a function of temperature. This work establishes a comprehensive fundamental framework for understanding spin relaxation/dissipation in GaAs-based FM/normal material (NM) heterostructures that may serve as the basis for coherent, time-resolved studies of spin transfer and dynamic exchange coupling in the emerging field of dynamically driven spin pumping. For example, while the current study focuses on the impact of the FPP process on the GaAs layer, symmetry argues that the exchange driven polarization of the photocarriers in GaAs must be accompanied by a concurrent *depolarization* of the Fe layer.

This manuscript is based upon work supported by the U.S. Department of Energy, Office of Science, Office of Basic Energy Sciences, under Award Number DE-FG02-03ER46054 and by the Center for Emergent Materials: an NSF MRSEC under award number DMR-1420451. The authors acknowledge John Carlin for the growth of the GaAs epitaxial layers, Mark Brenner for the deposition of the As capping layers, Kurtis Wickey for technical assistance and the NanoSystems Laboratory at the Ohio State University.

**FIG. 1.** (a) Top left panel: the spatial distribution of silicon donors and the inhomogeneous nuclear field resulting from electron spins trapped at the donor sites which hyperpolarize the surrounding nuclei at low temperature, as shown in the bottom left panel, which is the schematic potential profile along the white dashed line. Top right panel: a homogeneous nuclear field distribution at high temperature due to the delocalization of trapped carriers (bottom right panel). (b) Schematic of sample structure and time-resolved Kerr rotation (TRKR) measurement geometry. (c) Simulated band structure for sample in (b). Inset: calculated band structure near the GaAs/MgO/Fe interface showing that the Fermi level is pinned at 0.3 eV above the GaAs valence band maximum.

**FIG. 2.** (a) Measured Kerr rotation ($\theta_K$) vs $\Delta t$ for a Fe/MgO/GaAs heterostructure (solid circles) and a control GaAs epilayer (open circles) at $T=$ 5 K and $\boldsymbol{B}_{app}=$ 12 kG. The data are offset for clarity. (b) A cartoon illustrates that a nuclear field antiparallel to the applied field in a Fe/MgO/GaAs heterostructure results from the hyperfine coupling between GaAs nuclear spins ($\boldsymbol{I}$), and two non-equilibrium spin populations, $\boldsymbol{S}_{rel}$ and $\boldsymbol{S}_{FPP}$ (see text). (c) Top panel: total field $B_{tot}$ (Larmor frequency $\omega_L$) as a function of $B_{app}$ between -5 kG and +5 kG for Fe/MgO/GaAs and bare GaAs at $T=$ 5 K. Bottom panel: nuclear field $B_n^{tot}$ ($B_n^{tot} = B_{tot} - B_{app}$) as a function of $B_{app}$.

**FIG. 3.** (a) Spin relaxation time ($T_2^*$, solid diamonds) and (b) $|B_n^{tot}|$ (open circles) and $1/T_2^*$ (solid diamonds) as a function of $B_{app}$ for Fe/MgO/GaAs up to 20 kG at $T=$ 5 K. (c) (d) Same data set as (a) and (b) with field up to 4 kG.

**FIG. 4.** (a) $|B_n^{tot}|$ and (b) $T_2^*$ as a function of temperature for $B_{app}$= 0.18 kG (black circles) and 0.10 kG (red circles). The black dashed line is the DP prediction of the temperature dependence of $T_2^*$.

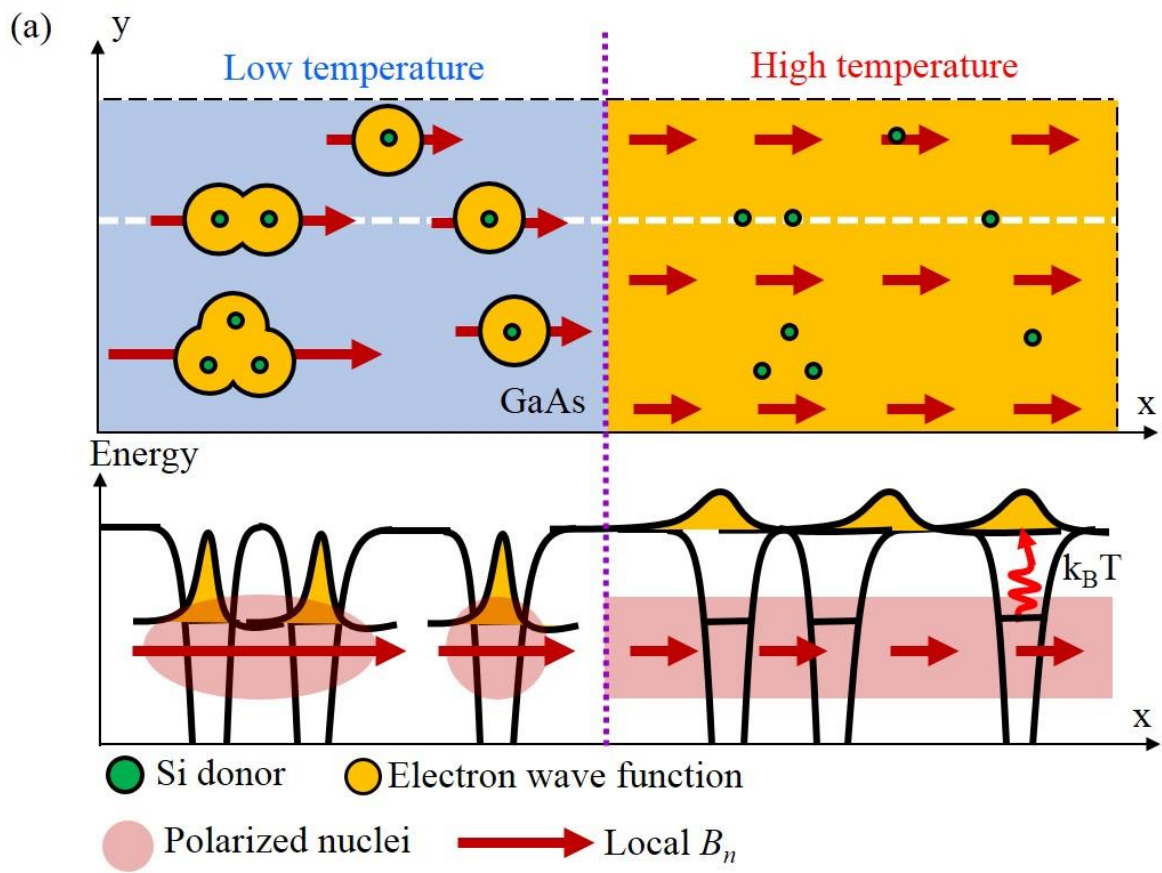

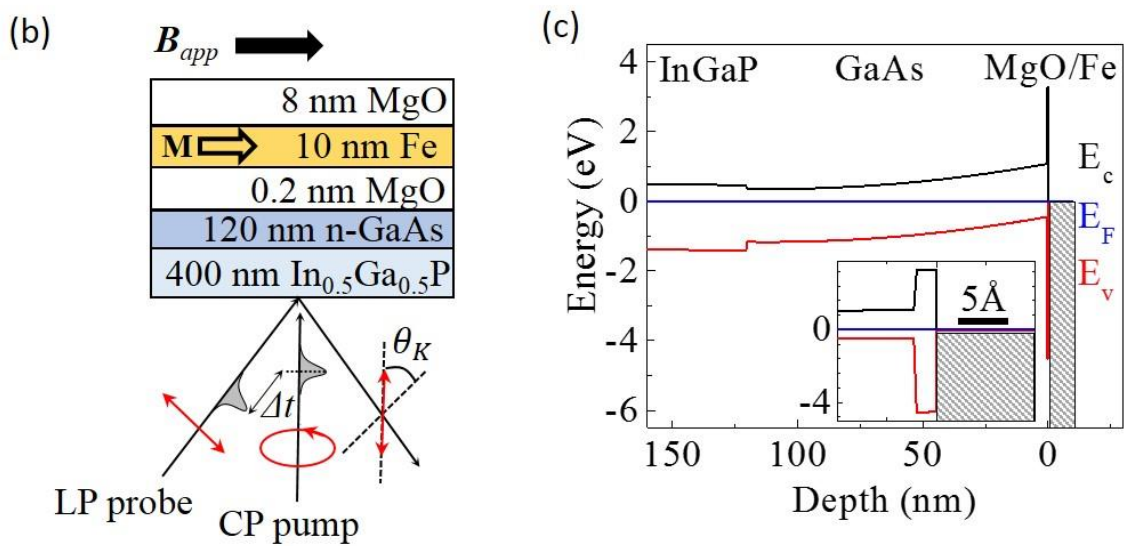

**Figure 1** Yu-Sheng Ou

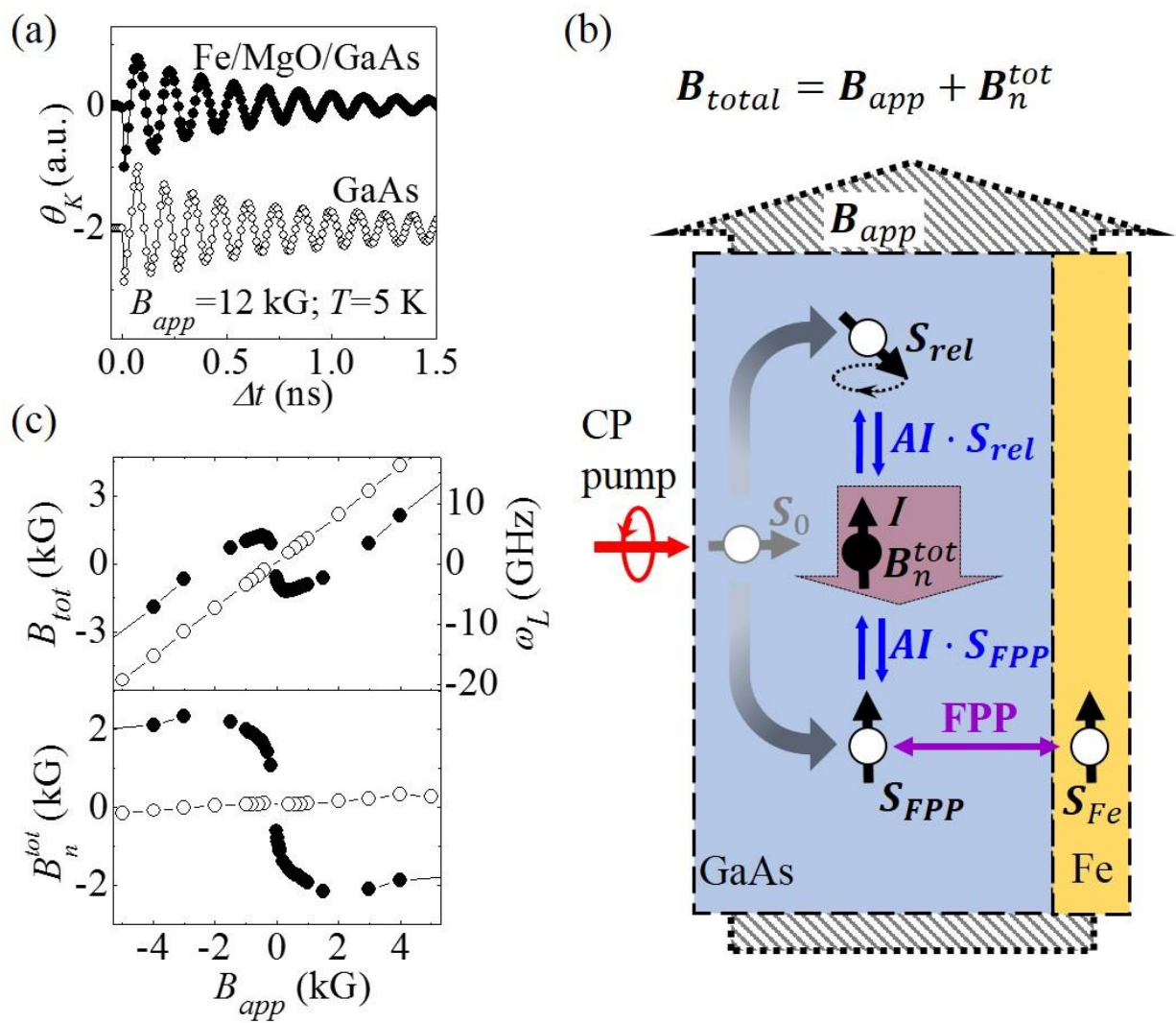

**Figure 2** Yu-Sheng Ou

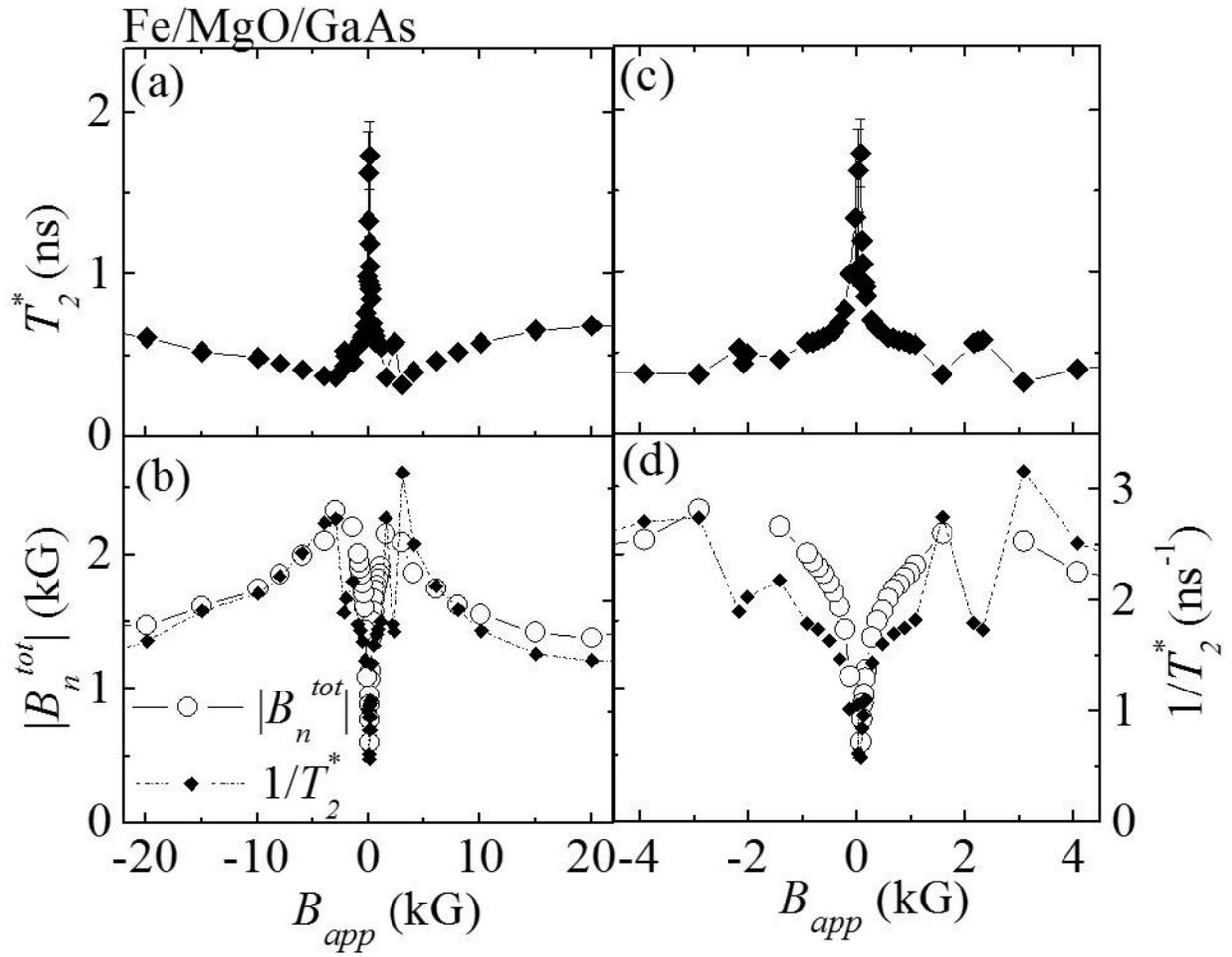

**Figure 3** Yu-Sheng Ou

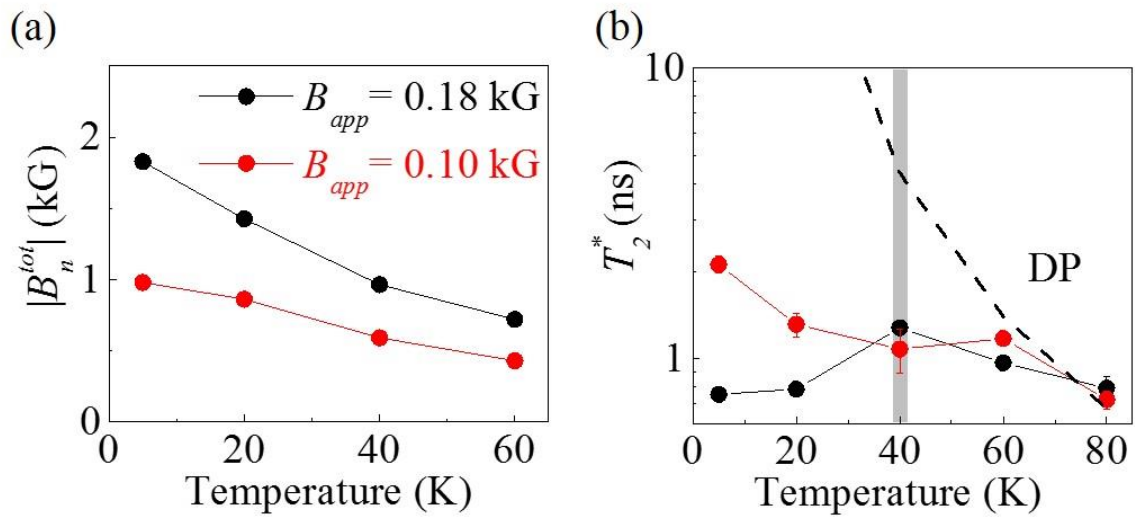

**Figure 4** Yu-Sheng Ou

*Supplemental materials*

**Exchange-driven spin relaxation in ferromagnet/oxide/semiconductor heterostructures**

Yu-Sheng Ou[1], Yi-Hsin Chiu[1], N. J. Harmon[2], Patrick Odenthal[3], Matthew Sheffield[1], Michael Chilcote[1], R. K. Kawakami[1,3], M. E. Flatté[2], E. Johnston-Halperin[1†]

[1]Department of Physics, The Ohio State University, Columbus, OH 43210-1117, USA
[2]Department of Physics & Astronomy, The University of Iowa, Iowa City, IA 52242-1479, USA
[3]Department of Physics & Astronomy, University of California, Riverside, CA 92521, USA

**A. All-optical nuclear magnetic resonance (NMR) to verify the nuclear origin of local field in Fe/MgO/GaAs heterostructures**

The premise that the difference between the local field measured by the Larmor precession of the electrons in Fig. 2 and the applied field is due to hyperfine coupling is supported by the temperature dependence of the local field (Fig. 1(a)) and the lab-time dependence of the TRKR signal (the signal evolves on the scale of minutes when the experimental parameters are changed). This hypothesis can be further validated by exploring the resonant depolarization of the presumed nuclear spin orientation using the repetition rate of the pulsed laser (76 MHz) as a periodic tipping pulse [36, S1, S2]. This depolarization can be driven by two processes: (1) the electron spins excited by periodic laser pump pulses generate an effective periodic hyperfine field that resonantly depolarizes a specific nuclear species at appropriate applied field; (2) periodically photoexcited carriers create a modulated electric field that in turn induces quadrupolar resonance of a specific nuclear species. The quadrupolar resonance in case (2) occurs at half of the applied field in case (1) because it involves a transition of $\Delta m = 2$ (m is nuclear spin quantum number) rather than a $\Delta m = 1$ transition, as in (1). A sensitive and less time-consuming approach for probing this optically-pumped NMR can be accomplished by measuring the Kerr rotation angle ($\theta_K$) as a function of applied field ($B_{app}$) at fixed delay time ($\Delta t$). As can be seen in Eq. (1) in the main text $\theta_K$ has the same functional dependence on $B_{tot}$ and $\Delta t$, so in the absence of nuclear effects (when $B_{tot} = B_{app}$) varying $B_{app}$ should have the same effect as varying $\Delta t$, i.e. an oscillating cosine with frequency given by $g\mu_B \Delta t/\hbar$. When nuclear effects are included, the additional hyperfine field, $B_n$, results in a phase shift of these oscillations off-resonance that is abruptly suppressed with $B_{app}$ matches the resonance condition for one of the nuclear spin species.

The top panel of Fig. S1 shows $\theta_K$ as a function of $B_{app}$ at $\Delta t = 400$ ps and $T = 5$ K. To clearly resolve the NMR features, the oscillatory background is subtracted, as shown in the bottom panel. Four resonance peaks at $B_{app}=$ 59.1 kG, 52.4 kG, 37.5 kG and 29.5 kG correspond to nuclear dipole resonance of $^{71}$Ga ($\gamma = 1.02475$ MHz/kG), and nuclear quadrupolar resonances of $^{75}$As ($\gamma = 0.73148$ MHz/kG), $^{69}$Ga ($\gamma = 1.30204$ MHz/kG) and $^{71}$Ga respectively. These results are consistent with previous all-optical NMR studies on bulk GaAs [36] and GaAs quantum wells [S1, S2], and confirm the nuclear origin of the local field in Fe/MgO/GaAs heterostructures.

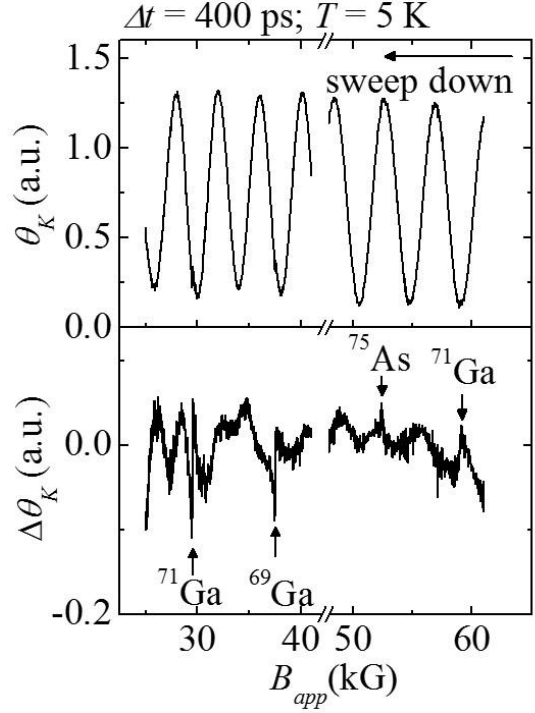

**Figure S1**: All-optical NMR study on Fe/MgO/GaAs heterostructure. Top panel: $\theta_K$ as a function of $\boldsymbol{B}_{app}$ at $\Delta t$ = 400 ps and $T$ = 5 K. Bottom panel: The same data set as the top panel, but the oscillatory background is subtracted for resolving small NMR features.

**B. Quantitative description of inhomogeneous nuclear field model and comparison with experimental data**

Our theoretical approach begins with the following equation for the evolution of a spin:

$$\frac{d\boldsymbol{S}(t)}{dt} = \gamma[\boldsymbol{B}_{app} + \boldsymbol{B}_n(t)] \times \boldsymbol{S}(t) \quad (S1)$$

Where $\gamma$ is the GaAs electron gyromagnetic ratio, the applied field, $\boldsymbol{B}_{app}$, is time-independent and the nuclear field, $\boldsymbol{B}_n$, possesses time dependence only from the point of view that an electron experiences it intermittently. An ensemble of spins randomly walks in between the different field environments as described in Ref. [9]. The final result yields an anisotropic spin relaxation term in our situation:

$$\frac{d\mathbf{S}(t)}{dt} = -\frac{\gamma^2}{4}\frac{B_n^2}{k_0}\mathbf{S}(t) \quad (S2)$$

Where we assume that the spin spends much more time in the nuclear field than outside the nuclear field. $k_0$ is the rate at which the spin enters the nuclear region. We have also assumed that the nuclear field is nearly collinear with the applied field. The spin relaxation rate can be derived from (S2) as $\frac{1}{T_2^*} = \frac{\gamma^2}{4}\frac{B_n^2}{k_0}$, which is used to fit the experimentally extracted $1/T_2^*$ verse $B_n^{tot}$ at $B_{app}$ between 0.1 kG and 0.28 kG and at $T= 5K$ as shown in Fig. S2. From the fitting, the extracted $k_0$ is 0.014 ps$^{-1}$, and $1/k_0$, which can be treated as the time it takes from one nuclear region (or donor site) to the other, is about 70ps.

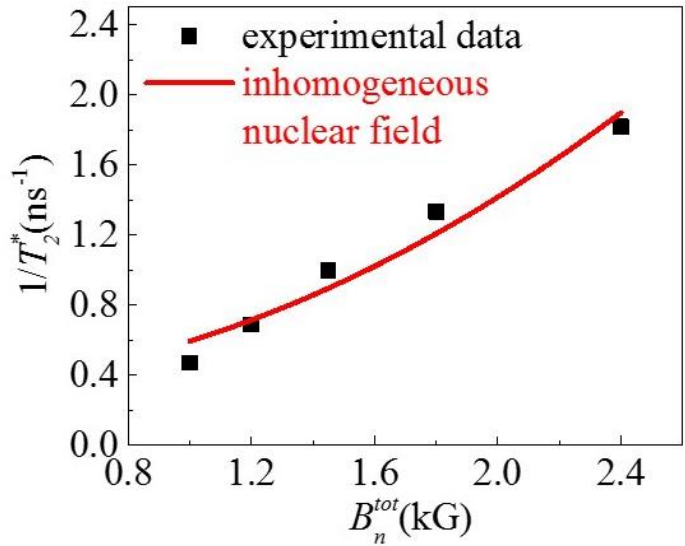

**Figure S2**: Spin relaxation rate ($1/T_2^*$) verse nuclear field strength ($B_n^{tot}$). Solid square represents our experimental data at $T= 5K$, and the red line is the fitting curve generated by the spin relaxation rate due to the inhomogeneity of nuclear field.

## C. Quantitative comparison with spin-phonon coupling spin relaxation mechanism at low applied field

In Ref. [7], the spin phonon coupling spin relaxation rate ($\tau_{s-ph}$) is derived as

$$\frac{1}{\tau_{s-ph}} = \alpha_{s-ph} B_{app}^2 T^4 f(T) \quad (S3)$$

Where $f(T) = \int_0^{\frac{343}{T}} x^3 \left[\frac{1}{2} + \frac{1}{e^x - 1}\right] dx$ and $\alpha_{s-ph} = 2.2 \times 10^{-13}$ kG$^{-2}$ K$^{-4}$ ns$^{-1}$. The calculated $1/\tau_{s-ph}$ at $T=$ 5K and $B_{app}$ between 0.1 kG and 0.28 kG is compared to our data as shown in Fig. S3. Within this field region, the calculated $1/\tau_{s-ph}$ is five orders of magnitude smaller than our data, and therefore can't to used explain the low temperature behavior of $T_2^*$.

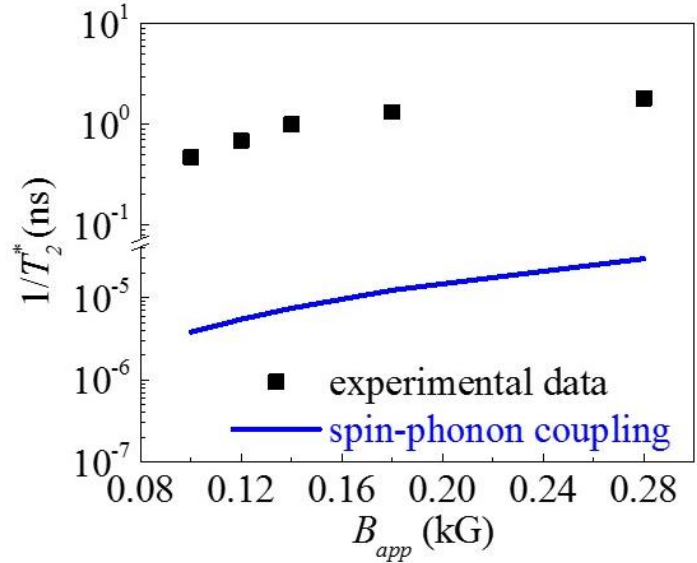

**Figure S3**: Spin relaxation rate ($1/T_2^*$) verse applied field ($B_{app}$). Solid squares represents our experimental data at $T=$ 5K, and the blue line is the calculated spin relaxation rate due to spin-phonon coupling.

**References:**

[S1] G. Salis, D. T. Fuchs, J. M. Kikkawa, D. D. Awschalom, Y. Ohno, and H. Ohno, Phys. Rev. Lett. **86**, 2677 (2001).

[S2] G. Salis, D. D. Awschalom, Y. Ohno, and H. Ohno, Phys. Rev. B **64**, 195304 (2001).